\documentclass[11pt,a4paper]{article}

\usepackage[margin=1.05in]{geometry}
\usepackage[T1]{fontenc}
\usepackage{lmodern}
\usepackage{graphicx}
\usepackage{amsmath}
\usepackage{booktabs}
\usepackage{array}
\usepackage{url}
\usepackage{multirow}
\usepackage[hidelinks]{hyperref}
\usepackage{xurl}
\usepackage[font=small,labelfont=bf]{caption}
\usepackage{microtype}
\usepackage{etoolbox}
\apptocmd{\thebibliography}{\small\setlength{\baselineskip}{11.5pt}\setlength{\itemsep}{0pt}\setlength{\parsep}{0pt}}{}{}

\graphicspath{{figures/}}
\hypersetup{
  pdftitle={Visual Orchestration Tax in Agentic VLM Pipelines: Auditing and Certifying Visual Evidence Reuse},
  pdfauthor={Lingteng Zeng},
  pdfsubject={Visual evidence reuse in agentic vision-language pipelines},
  pdfkeywords={multi-agent systems, vision-language models, visual evidence reuse, orchestration tax, redundancy audit}
}
\newlength{\positioningwidth}

\title{Visual Orchestration Tax in Agentic VLM Pipelines:\\
Auditing and Certifying Visual Evidence Reuse}
\author{Lingteng Zeng\\[0.35em]
  \small Faculty of Engineering, The Chinese University of Hong Kong\\
  \small Hong Kong, China\\
  \small \href{mailto:lingtengzeng@link.cuhk.edu.hk}{\texttt{lingtengzeng@link.cuhk.edu.hk}}\\
  \small \href{https://orcid.org/0009-0006-0362-630X}{\raisebox{-0.15em}{\includegraphics[height=1em]{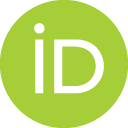}}\hspace{0.5em}\texttt{https://orcid.org/0009-0006-0362-630X}}}
\date{}

\begin{document}
\maketitle

\begin{abstract}
Agentic VLM pipelines increasingly pass the same static visual evidence through multiple specialist agents and tools.
This design creates an orchestration-level redundancy mode: semantically unchanged images are repeatedly reconstructed as image-conditioned requests at the VLM API boundary.
We call this phenomenon \emph{visual orchestration tax} and develop a measurement-to-certification framework for visual evidence reuse in agentic VLM pipelines.
The audit side defines $\mathrm{M1}_{\mathrm{trace}}$ to count raw visual-evidence touches and M2 to measure structural touch redundancy, with query-level distributions, bootstrap confidence intervals, and paired quality tests.
Across SeeingEye and MAMMQA on chart, document, general-VQA, and multi-modal-QA tasks, audits reveal 66.8--75.6\% visual-evidence touch redundancy, and every audited query exceeds the predefined gate.
The certification side introduces \textsc{SharedVisCache}, a contract-aware evidence reuse hook keyed by image content, preprocessing fingerprint, and encoder assumptions.
On SeeingEye, contract validation certifies 75.0--75.5\% repeated touches as reusable while preserving 350/350 output strings and $\Delta\mathrm{M5}{=}0$.
At the physical layer, certified hits reduce $F_{\mathrm{vision}}$ from 800 to 200 in ChartQA-200 trace replay and from 200 to 50 inside live SeeingEye translator-stage physical integration, preserving 800/800 replay strings and 200/200 integrated call outputs.
The results position visual reuse as a measurable, behavior-preserving property of agent orchestration and define an agent-layer contract that makes backend prefix or token reuse semantically interpretable.
\end{abstract}


\section{Introduction}
\label{sec:intro}

Human--AI and multi-agent vision-language (VLM) systems increasingly decompose visual question answering into specialist modules---translators, reasoners, and modality experts~\cite{seeingeye2025,mammqa2025,gamagent2025}.
This modularity clarifies reasoning, but it also creates an orchestration problem: the same static image can be re-attached at multiple agent/tool turns, repeatedly exposing unchanged visual evidence to the VLM API.
We call this agent-layer redundancy \emph{visual orchestration tax}.
Benchmark accuracy and serving-engine counters leave this movement of visual evidence implicit, because the redundancy originates in how agent frameworks package multimodal context.
The central question is whether an agent pipeline repeatedly constructs semantically equivalent image-conditioned requests before any serving-layer optimization can be interpreted.
Thus, our principle is audit before acceleration and certify before reuse: static visual evidence must first be exposed as an agent-layer object, and reuse requires an explicit image-preprocessing-encoder contract that preserves behavior.

Serving-level reuse (e.g., VLCache~\cite{vlcache2025}) and text-side KV sharing across agents (KVCOMM~\cite{kvcomm2025}) optimize different layers of the stack.
Trained single-policy frameworks with native visual-prefix reuse (Visual Para-Thinker++~\cite{parathinker2026}) require retraining and homogeneous backbones.
\textbf{Existing work centers on backend artifacts, text-KV communication, or retrained policies; this paper audits the off-the-shelf agent orchestration layer and validates contract-certified visual reuse under paired equal-quality constraints.}

We study two questions.
\textbf{RQ1 (measurement)} asks how much visual-evidence touch redundancy arises in off-the-shelf agentic VLM pipelines and how it can be audited reproducibly.
\textbf{RQ2 (reuse correctness)} asks whether repeated image-conditioned touches can be certified as reusable under an explicit visual-evidence contract without perturbing paired task behavior or the agent reasoning path.

Our contributions follow this measurement-to-certification path:
\begin{enumerate}
  \item \textbf{Visual orchestration tax as a measurable redundancy mode.} We formalize repeated static visual-evidence reconstruction as an orchestration-layer redundancy mode and introduce $\mathrm{M1}_{\mathrm{trace}}$/M2 metrics with bootstrap confidence intervals and query-level robustness checks.
  \item \textbf{A contract for reusable visual evidence.} We introduce \textsc{SharedVisCache}, a contract-aware evidence reuse interface keyed by image content, preprocessing fingerprint, and encoder assumptions, turning visual reuse from an implicit backend side effect into an explicit agent-level property.
  \item \textbf{Paired reuse-correctness validation and physical realization.} Across two open agentic VLM pipelines and four visual domains, we observe 66.8\%--75.6\% visual-evidence touch redundancy. On SeeingEye, contract-level reuse certifies 75.0\%--75.5\% repeated touches as reusable while preserving 350/350 output strings and $\Delta$M5=0; trace-aligned replay and live translator-stage physical integration map the same contract hits to skipped Qwen2.5-VL vision-tower forwards.
\end{enumerate}
By placing this contract at the agent layer, the framework determines when visual reuse is semantically valid and behavior-preserving before backend-level prefix or token reuse is invoked.

\section{Related Work}
\label{sec:related}

\textbf{Multi-agent visual reasoning.}
Visual pipelines such as SeeingEye~\cite{seeingeye2025} alternate VLM perception with LLM reasoning, MAMMQA~\cite{mammqa2025} chains text, table, and image specialists, and GAM-Agent~\cite{gamagent2025} coordinates agents under uncertainty.
These designs improve interpretability and accuracy, but their efficiency profile is usually described by final-task accuracy and coarse runtime while the internal movement of visual evidence remains implicit.
Our audit makes this movement observable without changing the agent topology.

\textbf{Efficient VLM inference.}
Serving systems such as vLLM/PagedAttention optimize KV memory for high-throughput LLM serving~\cite{vllm2023}.
VLCache~\cite{vlcache2025} reuses vision-token KV across \emph{queries} at the serving engine; vLLM automatic prefix caching similarly targets repeated prompt-prefix work inside the serving stack~\cite{vllmapc2026}.
Quantization, batching, and native serving caches lower backend cost after requests are materialized; our unit of analysis is the visual evidence object repeatedly emitted by the agent loop at the API boundary.
This layer complements serving optimization by specifying which visual prefixes are semantically equivalent before backend reuse is invoked.

\textbf{Multi-agent LLM communication.}
KVComm~\cite{kvcomm2025} shares text KV across LLM agents and assumes vision was encoded once upstream.
SharedVisCache is complementary: it operates at the VLM API boundary before text-KV methods apply.

Table~\ref{tab:positioning} summarizes the resulting layering.
SharedVisCache applies at the API boundary and requires no retraining.

\begin{table}[t]
\caption{Layering of representative reuse mechanisms and complementary layers.}
\label{tab:positioning}
\centering
\footnotesize
\setlength{\positioningwidth}{\dimexpr\linewidth-8\tabcolsep\relax}
\begin{tabular}{@{}>{\raggedright\arraybackslash}p{0.15\positioningwidth}>{\raggedright\arraybackslash}p{0.22\positioningwidth}>{\raggedright\arraybackslash}p{0.17\positioningwidth}>{\raggedright\arraybackslash}p{0.24\positioningwidth}>{\raggedright\arraybackslash}p{0.22\positioningwidth}@{}}
\toprule
Mechanism & Reuse object & Layer & Validated effect & Complementary layer \\
\midrule
VLCache~\cite{vlcache2025} & vision-token KV & serving engine & backend vision-token reuse & agent trace redundancy \\
vLLM APC~\cite{vllmapc2026} & prompt/KV prefix & serving flag & shared-prefix prefill reuse & evidence-validity contract \\
KVComm~\cite{kvcomm2025} & text KV & agent communication & text-side KV sharing & visual evidence reconstruction \\
Para-Thinker++~\cite{parathinker2026} & trained visual policy & retraining / new policy & learned visual reasoning behavior & off-the-shelf API deployment \\
\textsc{SVC} (ours) & image + preprocess + encoder contract & VLM API / orchestration & paired reuse + VLM-boundary validation & serving-cache configuration \\
\bottomrule
\end{tabular}
\end{table}

\section{Redundancy Audit Protocol}
\label{sec:audit}

\subsection{Problem Setup and Metrics}

For each query $q$ with static image set $\mathcal{I}_q$, the audit metric is the raw trace count
\begin{equation}
  \mathrm{M1}_{\mathrm{trace}}(q) = \#\text{image-conditioned visual touches}, \quad
  \mathrm{M2}(q) = \frac{\mathrm{M1}_{\mathrm{trace}}(q) - |\mathcal{I}_q|}{\mathrm{M1}_{\mathrm{trace}}(q)} .
  \label{eq:m1m2}
\end{equation}
We report query means, medians, p90/max values, and non-parametric bootstrap 95\% confidence intervals for query means.
\textbf{Metric semantics.} M2 is a structural touch-redundancy statistic; compute, latency, and physical forward effects are reported through $V_{\mathrm{API}}$, $F_{\mathrm{vision}}$, and timed regions.
\textbf{Measurement gate:} $\mathrm{M2} \ge 0.25$.
For contract validation, \textbf{M5} denotes the benchmark-specific score (ChartQA relaxed accuracy, MMMU accuracy, and DocVQA ANLS), with a gate of $|\Delta\mathrm{M5}|{\le}0.01$ plus at least 30\% reduction from baseline $\mathrm{M1}_{\mathrm{trace}}$ to SharedVisCache $\mathrm{M1}_{\mathrm{contract}}$.
For paired quality, we report exact McNemar tables and exact 95\% upper bounds on the unseen discordance rate when no discordant pair is observed.
We report $V_{\mathrm{API}}$ separately for VLM server calls and $F_{\mathrm{vision}}$ for instrumented physical vision-tower forward calls in physical replay and integrated mechanism validation.
Tables abbreviate $\mathrm{M1}_{\mathrm{trace}}$ as $\mathrm{M1}_{t}$ and $\mathrm{M1}_{\mathrm{contract}}$ as $\mathrm{M1}_{c}$.

\paragraph{Evidence layers.}
We separate four evidence layers: trace ($\mathrm{M1}_{\mathrm{trace}}$/M2), contract (image-preprocessing-encoder hits), behavior (paired identity, $\Delta$M5, McNemar tables, and discordance bounds), and serving ($V_{\mathrm{API}}$, repeated-prefix probes, and $F_{\mathrm{vision}}$).
This separation validates the agent-layer reuse contract before binding it to a particular serving-engine realization.

\subsection{Instrumentation}

\begin{figure}[t]
\centering
\includegraphics[width=0.94\linewidth]{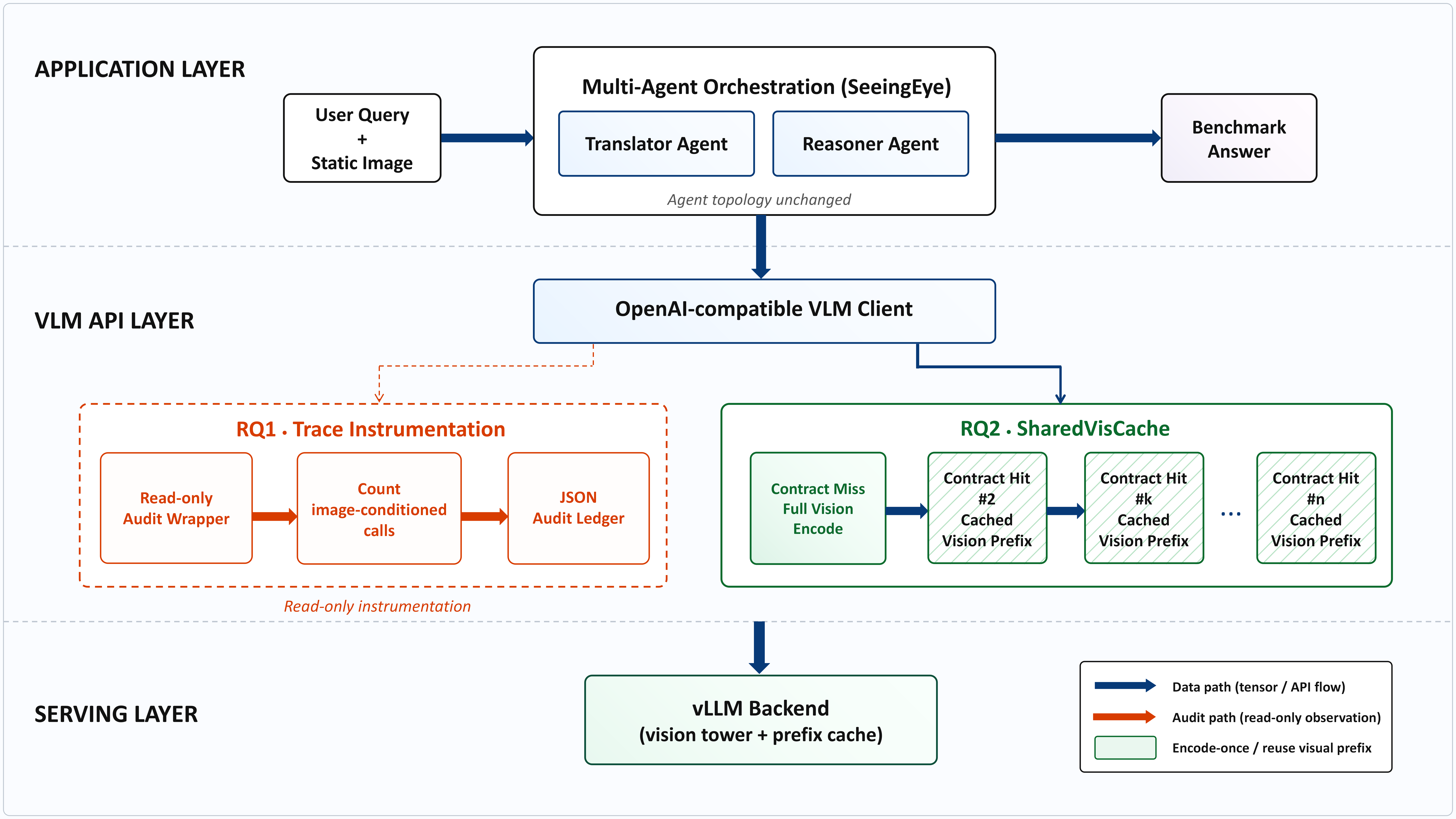}
\caption{Overall framework: redundancy audit instrumentation (RQ1) and contract-level reuse validation through SharedVisCache at the VLM API layer (RQ2). Baseline agent topology is unchanged.}
\label{fig:pipeline}
\end{figure}

Figure~\ref{fig:pipeline} shows the audit path.
We wrap the OpenAI-compatible VLM client and count messages containing \texttt{image\_url} blocks that carry visual evidence at the API boundary.
Unique images $|\mathcal{I}_q|$ are deduplicated by content hash.
Audit hooks observe passively; baseline agent logic and prompts are unchanged.
SeeingEye audits use phase-batched VLM-then-LLM execution; MAMMQA uses on-demand execution.

\paragraph{Worked example.}
On ChartQA, SeeingEye's translator issues four visual calls per query in the default inner tool loop, so $\mathrm{M1}_{t}{=}4$ and $\mathrm{M2}{=}0.75$.
The mean $\mathrm{M1}_{t}{=}4.00$ over $n{=}200$ (Table~\ref{tab:audit}) confirms this pattern at scale.

\subsection{Audit Results}

Table~\ref{tab:audit} reports live vLLM audits.
All rows exceed the measurement gate.

\begin{table}[t]
\caption{Visual-evidence touch redundancy audit (RQ1). Brackets are bootstrap 95\% CIs for query means. Live vLLM; sequential single-instance serving.}
\label{tab:audit}
\centering
\small
\begin{tabular}{llrccc}
\toprule
Pipeline & Benchmark & $n$ & $\mathrm{M1}_{t}$ mean [95\% CI] & $\mathrm{M1}_{t}$ p90/max & M2 mean [95\% CI] \\
\midrule
SeeingEye & MMMU-dev & 100 & 4.28 [4.04, 4.60] & 4/12 & 75.6 [75.1, 76.3] \\
SeeingEye & ChartQA test & 200 & 4.00 [4.00, 4.00] & 4/4 & 75.0 [75.0, 75.0] \\
SeeingEye & DocVQA val & 50 & 4.00 [4.00, 4.00] & 4/4 & 75.0 [75.0, 75.0] \\
MAMMQA & MMQA-dev & 100 & 28.44 [25.89, 30.93] & 45/45 & 66.8 [66.7, 67.0] \\
\bottomrule
\end{tabular}
\end{table}

SeeingEye exhibits $\approx$4 image-conditioned touches per static image on chart, document, and general-VQA subsets.
For MAMMQA, $\mathrm{M1}_{t}$ aggregates image-conditioned calls across three VLM stages and multi-image items; M2 is computed against unique images per query~\cite{mammqa2025}.
The distributional view avoids mean-level compression: ChartQA and DocVQA are fixed at four image-conditioned calls per query, while MMMU has a small tail (max 12) and MAMMQA has a larger multi-image tail (p90=45).
Trace-level robustness analysis further shows that 100\% of audited queries exceed the measurement gate, with minimum per-query M2 of 66.7\% across all four audit rows.
The top 10\% of queries account for 10.0--17.7\% of cacheable repeated touches, so the redundancy signal is distributed across the trace instead of concentrated in a tiny set of outliers.
A trace replay on MAMMQA gives a counterfactual exact-image reuse upper bound: 2844 image-conditioned touches contain 943 unique images and 1901 repeated-image touches (66.8\%) that are cacheable at the visual-evidence level.
SeeingEye and MAMMQA therefore play complementary roles: SeeingEye provides a controlled repeated-image inner loop for paired reuse-correctness validation, while MAMMQA provides an external multi-stage, multi-image trace stress test for visual orchestration tax.

\section{SharedVisCache}
\label{sec:cache}

\subsection{Design}

\begin{figure}[t]
\centering
\includegraphics[width=0.94\linewidth]{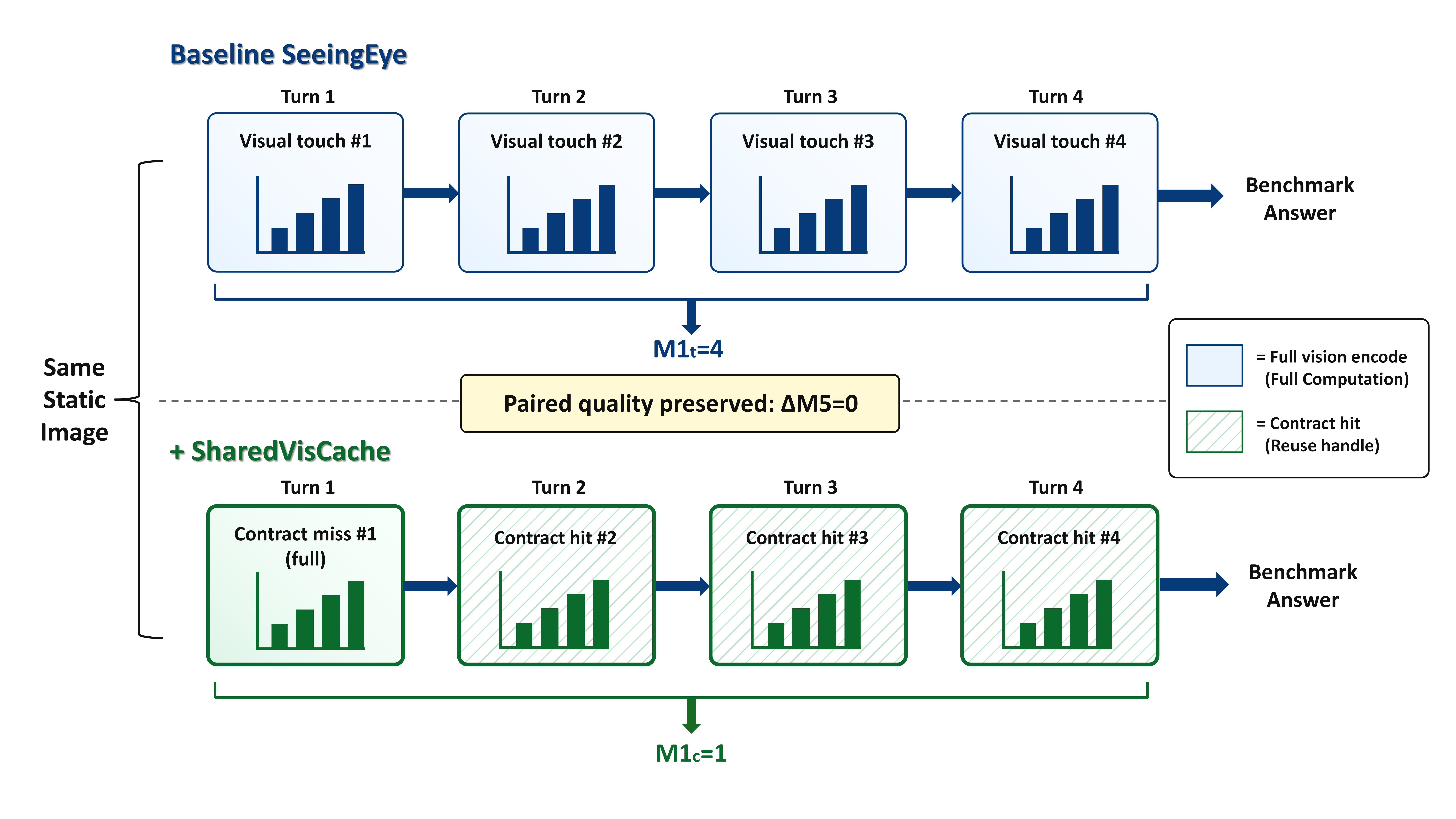}
\caption{Contract trace on a single static image within one SeeingEye query. SharedVisCache maps $\mathrm{M1}_{t}{=}4$ raw touches to $\mathrm{M1}_{c}{=}1$ certified evidence object while preserving paired answers; $V_{\mathrm{API}}$ and physical bypass are reported separately.}
\label{fig:encode_trace}
\end{figure}

SharedVisCache wraps the VLM API without forking baseline code (Fig.~\ref{fig:encode_trace}).
It makes repeated static visual evidence reusable at the orchestration boundary while holding agent prompts, tool selection, and answer generation fixed, which isolates the variable changed in paired experiments.
\paragraph{Definition 1.}
Two touches are contract-equivalent when their image hash and preprocessing fingerprint match under a declared encoder contract.
\textbf{Contract invariant:} under an unchanged prompt/tool trajectory, later equivalent touches may be replaced by contract hits while preserving visual-evidence identity.
\textbf{Certification scope.} Certification denotes implementation-level verification of contract equivalence and paired behavior in recorded executions.
\textbf{Cache key.} Identical images are keyed by a hash over image content and a declared preprocessing fingerprint, including resize, crop, normalization, model processor, and processor version assumptions.
The encoder contract states the visual-encoder path; heterogeneous revisions, adapters, quantization modes, dtypes, stochastic preprocessing, or image ordering require distinct declarations.
The publication runs use a fixed default fingerprint; if preprocessing changes, the cache key changes and reuse is rejected unless the new preprocessing is explicitly canonicalized.
\textbf{Contract-level accounting.} The hook records one contract miss for the first image-conditioned call and cache hits for later equivalent touches; $\mathrm{M1}_{c}$ is the contract-certified visual-evidence object count.
\textbf{API-compatible contract path.} Requests remain API-compatible with the VLM server and can exploit serving prefix-cache paths where available.
This path keeps server requests observable: Table~\ref{tab:cache} reports $\mathrm{M1}_{c}$ with $V_{\mathrm{API}}$, while Table~\ref{tab:physical} tests whether the same contract skips physical vision-tower forwards.

\subsection{Cacheability Contract}

The module preserves SeeingEye prompts, tool schemas, and agent topology.
The transport layer for vision evidence changes, enabling the same audit-and-reuse interface on OpenAI-compatible VLM backends.
Four cacheability controls validate the contract-aware key: identical image/same preprocessing hits, identical image/different question hits for the vision cache, different image/same question misses, and identical image/changed preprocessing misses.
Thus reuse requires matching image content and preprocessing fingerprint.
The contract is intentionally conservative.
Question text is excluded from the vision-cache key because the evaluated Qwen2.5-VL visual encoder path consumes the image independently of the natural-language question; query-conditioned visual selection or resampling belongs inside the encoder contract.
Preprocessing state is included because resize, crop, or normalization changes the visual token sequence.
This distinction turns cache safety from an implicit implementation assumption into an explicit interface property.
Equivalently, reuse is admitted under the contract
\[
  \mathrm{key}=\mathrm{image\_hash}+\mathrm{preprocess\_fingerprint}+\mathrm{encoder\_contract}.
\]

\section{Experiments}
\label{sec:experiments}

\subsection{Datasets and Setup}

\textbf{Hardware.} A single NVIDIA RTX 5090 accelerator (32\,GB).

\textbf{Models.} The VLM is Qwen2.5-VL-3B-Instruct; the LLM is Qwen3-8B.
Both are served through vLLM 0.23.0 with FlashInfer sampler disabled.
Server logs record bf16 execution, maximum model lengths of 16,384 (VLM) and 8,192 (LLM), GPU memory utilization 0.85, and prefix caching enabled unless ablated.

\textbf{Serving context.} The experiments use sequential single-instance serving, with one model resident on the accelerator at a time.
VLM$\leftrightarrow$LLM model-swap overhead is logged separately from $\mathrm{M1}_{t}$/M2 and contract-level counters.

\textbf{Baselines.} SeeingEye~\cite{seeingeye2025} for paired reuse evaluation; MAMMQA~\cite{mammqa2025} for audit and trace replay.

\textbf{Benchmarks.} ChartQA~\cite{chartqa2022}, MMMU-dev~\cite{mmmu2024}, MMQA-dev~\cite{talmor2021mmqa}, DocVQA validation~\cite{docvqa2021} ($n{=}50$ audit).

\subsection{Evaluation Protocol}

For RQ2 we run \emph{paired} same-session baseline vs.\ +SharedVisCache with shared weights and decoding configuration.
Acceptance follows predefined gates: M5 with $|\Delta\mathrm{M5}|{\le}0.01$, and at least 30\% reduction from baseline $\mathrm{M1}_{t}$ to SharedVisCache $\mathrm{M1}_{c}$.
The contract-level reuse factor is the baseline $\mathrm{M1}_{t}$ to SharedVisCache $\mathrm{M1}_{c}$ ratio, which is 4.0$\times$ on ChartQA subsets.
Paired output-string identity is 350/350 across all reported cache subsets.
We additionally report actual cache hit rates, paired exact McNemar tests, and exact upper bounds on correctness-discordance rates.
All publication numbers are drawn from machine-readable result ledgers and cross-checked by an automated consistency script.
The artifact ledger stores per-query JSONL rows with question/image IDs, answer strings, $\mathrm{M1}_{c}$, $V_{\mathrm{API}}$, cache hits/misses, wall-time fields, trace schema, audit wrappers, run configurations, and result ledgers.
Auxiliary trace analyses derive from the same traces without additional model answers.

\subsection{Redundancy Audit (RQ1)}

Table~\ref{tab:audit} summarizes structural redundancy.
M2 ranges from 66.8\% to 75.6\% across chart, document, general VQA, and multi-modal QA audits, indicating repeated reconstruction of identical static visual evidence in the audited modular pipelines.

\subsection{Contract-Level Reuse Validation (RQ2)}

Table~\ref{tab:cache} reports paired contract validation on SeeingEye; every row passes the reuse gate.
Scores reflect the 3B/8B model pair, and interpretation uses paired $\Delta$M5 plus the transition from baseline $\mathrm{M1}_{t}$ to SharedVisCache $\mathrm{M1}_{c}$.
The table also reports exact output-string identity and $V_{\mathrm{API}}$ to separate behavioral equivalence, contract-level reuse, and serving-call counters.
For DocVQA, M5 is ANLS.
Exact match (54.0\%/54.0\%) supplies the paired discordance table.

\begin{table}[t]
\caption{Contract-level reuse validation on SeeingEye. The third column maps baseline trace touches $\mathrm{M1}_{t}$(B) to SharedVisCache contract-certified objects $\mathrm{M1}_{c}$(SVC), while $V_{\mathrm{API}}$ reports the VLM serving-call counter. ChartQA uses relaxed accuracy, MMMU uses accuracy, and DocVQA uses ANLS. No correctness-discordant pairs were observed, so McNemar exact $p{=}1.000$ for each row and the last column reports an exact 95\% upper bound on discordance.}
\label{tab:cache}
\centering
\scriptsize
\begin{tabular}{lrrrrrrr}
\toprule
Benchmark & $n$ & $\mathrm{M1}_{t}\!\to\!\mathrm{M1}_{c}$ & $V_{\mathrm{API}}$ B/SVC & Hit (\%) & M5 B/SVC & Exact out & Disc.\ UB (\%) \\
\midrule
ChartQA (rel. acc.) & 50  & 4.0$\to$1.0 & 4.0/4.0 & 75.0 & 60.0/60.0 & 50/50 & 5.82 \\
ChartQA (rel. acc.) & 200 & 4.0$\to$1.0 & 4.0/4.0 & 75.0 & 54.0/54.0 & 200/200 & 1.49 \\
MMMU-dev (acc.) & 50 & 4.08$\to$1.0 & 4.08/4.08 & 75.5 & 28.0/28.0 & 50/50 & 5.82 \\
DocVQA (ANLS) & 50 & 4.0$\to$1.0 & 4.0/4.0 & 75.0 & 60.4/60.4 & 50/50 & 5.82 \\
\bottomrule
\end{tabular}
\end{table}

Across the 350 paired examples, baseline-correct/cache-wrong and cache-correct/baseline-wrong changes are both zero; pooled descriptively, the exact 95\% upper bound on unseen discordance is 0.85\%.

\subsection{Ablation Studies}

Table~\ref{tab:ablation} tests whether redundancy and contract-level reuse gains are robust to outer-loop depth and the prefix-caching serving knob.

\begin{table}[t]
\caption{Ablation on ChartQA-50 (SeeingEye). Iteration rows report audit M2; SVC rows report baseline $\mathrm{M1}_{t}$ to SharedVisCache $\mathrm{M1}_{c}$ reuse.}
\label{tab:ablation}
\centering
\small
\begin{tabular}{lrrr}
\toprule
Setting & Count & M2 / Reuse (\%) & M5 (B / SVC) \\
\midrule
\texttt{max\_iterations}${=}1$ & $\mathrm{M1}_{t}$ 4.0 & 75.0 & --- \\
\texttt{max\_iterations}${=}2$ & $\mathrm{M1}_{t}$ 4.0 & 75.0 & --- \\
\texttt{max\_iterations}${=}3$ & $\mathrm{M1}_{t}$ 4.0 & 75.0 & --- \\
SVC validation (prefix on) & $\mathrm{M1}_{t}\!\to\!\mathrm{M1}_{c}$ 4.0$\to$1.0 & 75.0 & 60.0 / 60.0 \\
SVC validation (prefix off) & $\mathrm{M1}_{t}\!\to\!\mathrm{M1}_{c}$ 4.0$\to$1.0 & 75.0 & 60.0 / 60.0 \\
\bottomrule
\end{tabular}
\end{table}

\textbf{Outer-loop depth.}
Varying \texttt{max\_iterations} from 1 to 3 leaves $\mathrm{M1}_{t}{=}4.0$ and M2$=$75.0\% unchanged: redundancy is driven by the translator \emph{inner} tool loop, with the varied outer iteration count producing no change.

\textbf{Prefix caching.}
With prefix caching disabled, the $\mathrm{M1}_{t}\!\to\!\mathrm{M1}_{c}$ reduction and M5 remain unchanged (60.0\% / 60.0\%), indicating that this knob mainly affects serving-layer KV latency.

\subsection{Systems Implications and Reuse Envelope}

SharedVisCache certifies about 75\% of repeated image-conditioned touches as reusable with zero paired M5 drop across ChartQA, MMMU, and DocVQA subsets.
The ablation localizes this reuse to the translator inner loop; outer-loop depth and the prefix-caching flag leave the contract-level reuse rate unchanged.
Because sequential VLM/LLM residency dominates the deployment (33,703.7\,s aggregate model-swap time on ChartQA $n{=}200$), the $\mathrm{M1}_{t}\!\to\!\mathrm{M1}_{c}$ transition is the primary orchestration-level effect-size metric; $V_{\mathrm{API}}$ is reported separately in Table~\ref{tab:cache}.

Table~\ref{tab:physical} reports two physical validations at the Qwen2.5-VL vision-tower boundary: ChartQA-200 trace replay and a live ChartQA-50 translator loop whose intercepted VLM calls are served through direct Qwen2.5-VL inference.
The cache stores synchronized vision-forward output under the image, preprocessing, and encoder key; later equivalent touches reuse this object instead of invoking the original vision tower.
In the integrated run, SeeingEye still issues 200 VLM calls, but 150 contract hits skip physical vision-tower forwards while preserving 50/50 translator outputs and 200/200 per-call VLM outputs.
Timing covers the CUDA-synchronized vision-tower forward region, isolating physical realization from decoding and full agent scheduling.
Accordingly, the live row is interpreted as translator-stage VLM-boundary realization; decoder and scheduler effects remain outside the timed region.
Inputs with mismatched image content, preprocessing fingerprint, or encoder contract become forced misses; a 20-touch changed-preprocessing control produces 20 misses in replay.

\begin{table}[t]
\caption{Physical realization of SharedVisCache at the Qwen2.5-VL vision-tower boundary. Trace rows replay the ChartQA-200 validation rows from Table~\ref{tab:cache}; live rows run the SeeingEye translator loop on ChartQA-50 and intercept its OpenAI-compatible VLM calls. Timing reports CUDA-synchronized vision-tower forward calls. Invalid image, preprocessing, or encoder keys are forced misses by the contract.}
\label{tab:physical}
\centering
\scriptsize
\begin{tabular}{lrrrrrr}
\toprule
Setting & Queries & Touches & $F_{\mathrm{vision}}$ & Vision time & F/time red. & Output identity \\
\midrule
Trace full encode & 200 & 800 & 800 & 37.0803\,s & -- & -- \\
Trace physical SVC & 200 & 800 & 200 & 9.8358\,s & 75.0/73.5 & 800/800 \\
Live full encode & 50 & 200 & 200 & 10.0894\,s & -- & -- \\
Live physical SVC & 50 & 200 & 50 & 2.0148\,s & 75.0/80.0 & 200/200 \\
\bottomrule
\end{tabular}
\end{table}

As a serving-context probe separate from the primary mechanism evidence, a direct VLM microbenchmark on 90 repeated static images reports repeated-request TTFT ratios of 2.06$\times$/2.00$\times$ with vLLM prefix caching and 2.04$\times$/1.99$\times$ with prefix caching disabled.
A normalized lookup-overhead analysis bounds trace-level savings at 63.5--72.8\% with 5\% hit cost and 50.1--57.5\% with a conservative 25\% hit-cost assumption.

\paragraph{When the contract matters.}
The tax is largest when chart, document, or multi-image QA pipelines repeatedly ask agents or tools to inspect the same static visual evidence.
It is weaker for one-shot calls, dynamic streams, or heterogeneous preprocessing, where SharedVisCache misses unless preprocessing is explicitly canonicalized.

\section{Limitations and Future Work}
\label{sec:limitations}

The evaluation covers static-image queries, SeeingEye paired contract validation, translator-stage physical integration, and a 3B/8B model pair on subset splits ($n{=}50$--200).
MAMMQA provides audit and trace replay, while DocVQA contract validation uses $n{=}50$ under the same model stack.
Trace-aligned replay and live translator-stage integration validate reusable visual computation at the VLM boundary; end-to-end latency in the present deployment is dominated by sequential VLM/LLM server swaps.
The cacheability contract is empirical and implementation-level; it is distinct from a formal semantic proof across arbitrary VLM backbones.
Heterogeneous preprocessing requires compatible fingerprints or adapter-level canonicalization; otherwise SharedVisCache misses by design.
Future work includes heterogeneous VLM agents, video streams, colocated serving with native multimodal caches and stable media identifiers, lookup/memory/eviction overhead, cost-weighted traces, MAMMQA reuse validation, and preprocessing canonicalization.

\section{Conclusion}
\label{sec:conclusion}

We presented visual orchestration tax as a measurable redundancy mode in agentic VLM pipelines and quantified consistently high M2 across two open systems and multiple benchmarks.
SharedVisCache shows that repeated visual-evidence touches can be converted into certified contract hits while preserving paired task behavior.
The physical replay and live translator-stage physical integration further show that certified hits can map to skipped physical vision-tower forwards, linking reuse correctness to a backend realization path.
By separating trace observability, contract certification, behavior preservation, and physical mechanism validation, the framework defines an agent-layer contract that specifies when visual evidence is reusable before backend prefix or token reuse is attempted.

\section*{Competing Interests}
The authors have no competing interests to declare that are relevant to the content of this article.

\end{document}